\documentclass[useAMS,usenatbib]{mn2e}
\usepackage{graphicx}
\title[Cosmic-ray heating of the early Universe]{Preheating of the
  Universe by cosmic rays from primordial supernovae at the beginning
  of cosmic reionization}

\author[S. Sazonov et al.]{S. Sazonov$^{1,2}$\thanks{E-mail:
sazonov@iki.rssi.ru} and R. Sunyaev$^{3,1}$\\
$^{1}$Space Research Institute, Russian Academy of Sciences,
Profsoyuznaya 84/32, 117997 Moscow, Russia\\
$^{2}$Moscow Institute of Physics and Technology, Institutsky per. 9,
141700 Dolgoprudny, Russia\\
$^{3}$Max-Planck-Institut f\"ur Astrophysik,
Karl-Schwarzschild-Str. 1, 85740 Garching bei M\"unchen, Germany
}

\newcommand{\beq}{\begin{equation}}
\newcommand{\eeq}{\end{equation}}
\newcommand{\beqa}{\begin{eqnarray}}
\newcommand{\eeqa}{\end{eqnarray}}

\newcommand{\lesssim}{\hbox{\rlap{\hbox{\lower4pt\hbox{$\sim$}}}\hbox{$<$}}}
\newcommand{\gtrsim}{\mathrel{\hbox{\rlap{\hbox{\lower4pt\hbox{$\sim$}}}\hbox{$>$}}}} 

\newcommand{\msun}{M_\odot}
\newcommand{\mpr}{m_p}

\newcommand{\tcmb}{T_{\rm{CMB}}}
\newcommand{\mhalo}{M_{\rm{h}}}
\newcommand{\eb}{E_{\rm{b}}}

\newcommand{\tvir}{T_{\rm{vir}}}
\newcommand{\esn}{E_{\rm{SN}}}
\newcommand{\fsn}{f_{\rm{SN}}}
\newcommand{\rvir}{R_{\rm{vir}}}
\newcommand{\fgas}{f_{\rm{gas}}}
\newcommand{\nh}{n_{\rm{H}}}
\newcommand{\nnh}{n_{\rm{HI}}}

\newcommand{\vsh}{V_{\rm{sh}}}
\newcommand{\rsh}{R_{\rm{sh}}}
\newcommand{\bsh}{B_{\rm{sh}}}

\newcommand{\elecr}{E_{\rm{LECR}}}
\newcommand{\emin}{E_{\rm{min}}}
\newcommand{\emax}{E_{\rm{max}}}
\newcommand{\pmax}{p_{\rm{max}}}
\newcommand{\xe}{x_{\rm{e}}}
\newcommand{\eheat}{E_{\rm{heat}}}
\newcommand{\fheat}{f_{\rm{heat}}}
\newcommand{\thub}{t_{H}}
\newcommand{\tauloss}{\tau_{\rm{loss}}}

\newcommand{\ti}{T_{\rm{IGM}}}
\newcommand{\bi}{B_{\rm{IGM}}}
\newcommand{\mmin}{M_{\rm{min}}}
\newcommand{\mmax}{M_{\rm{max}}}
\newcommand{\nhalo}{n_{\rm{h}}}
\newcommand{\omegam}{\Omega_{\rm{m}}}
\newcommand{\omegab}{\Omega_{\rm{b}}}
\newcommand{\rhom}{\rho_{\rm{m}}}
\newcommand{\rhoc}{\rho_{\rm{c}}}

\newcommand{\rfree}{R_{\rm free}}
\newcommand{\rmf}{R_B}
\newcommand{\rxray}{R_{\rm{X-ray}}}
\newcommand{\mpath}{\langle\lambda\rangle}

\newcommand{\ma}{M_{\rm A}}
\newcommand{\va}{v_{\rm A}}
\newcommand{\ms}{M_{\rm s}}
\newcommand{\cs}{c_{\rm s}}

\begin{document}

\maketitle

\label{firstpage}

\begin{abstract}

The 21-cm signal from the cosmic reionization epoch can shed light on
the history of heating of the primordial intergalactic medium (IGM) at
$z\sim 30$--10. It has been suggested that X-rays from the first
accreting black holes could significantly heat the Universe at these 
early epochs. Here we propose another IGM heating mechanism associated with
the first stars. As known from previous work, the remnants of powerful
supernovae (SNe) ending the lives of massive  Population III stars
could readily expand out of their host dark matter minihalos into the
surrounding IGM, aided by the preceeding photoevaporation of the
halo's gas by the UV radiation from the progenitor star. We argue that
during the evolution of such a remnant a significant fraction of the
SN kinetic energy can be put into low-energy ($E\lesssim 30$~MeV)
cosmic rays that will eventually escape into the IGM. These
subrelativistic cosmic rays could propagate through the Universe and
heat the IGM by $\sim 10$--100~K by $z\sim 15$, before more powerful
reionization/heating mechanisms associated with the first galaxies and
quasars came into play. Future 21-cm observations could thus constrain
the energetics of the first supernovae and provide information on the
magnetic fields in the primordial IGM.
 
\end{abstract}

\begin{keywords}
dark ages, reionization, first stars -- cosmic rays -- supernovae:
general.  
\end{keywords}

\section{Introduction}
\label{s:intro}

Altough it is clear that the reionization of hydrogen in the Universe
was essentially complete by $z\sim 6$, the history of cosmic
reionization remains poorly known (see \citealt{barloe01,fanetal06}
for reviews). The spin-flip transition of HI, at an observed
wavelength of $21(1+z)$~cm, provides a potentially powerful and unique
tool for studying the corresponding early epochs observationally (e.g.,
\citealt{morwyi10}). The properties of the 21-cm signal crucially
depend on the deviation of the HI spin temperature from that of the
cosmic microwave background (CMB) as a function of redshift. The spin
temperature in turn is sensitive to the kinetic temperature of the
(nearly neutral) medium (hereafter referred to as the intergalactic
medium, IGM) filling the early Universe and thus depends on the 
efficiency of any gas heating mechanisms that might have been at play
(see \citealt{furetal06,priloe12} for detailed discussions of the
physics involved). 

It is widely accepted that cosmic reionization was mostly driven
by stellar UV radiation from the first galaxies. Reionization by stars
operates through seeding, growing and merging of HII bubbles in the
IGM, with the medium outside the ionized regions remaining cold and
neutral. However, it has been suggested
\citep{venetal01,madetal04,ricost04,miretal11,tanetal12} that during 
the early epoch of star formation there possibly existed X-ray
sources, such as microquasars and high-mass X-ray binaries, which
could significantly heat the gas throughout the Universe by their
radiation thanks to the low opacity of the IGM to X-rays compared to
UV radiation. As a result of such X-ray preheating, the spin
temperature of the HI gas could become coupled to its kinetic
temeperature already by $z\sim 15$. This would make the 21-cm signal
from the early phases of cosmic reionization quite different from what
it would be in the absence of such heating
\citep{prifur07,priloe10,mesetal13,fiaetal14}. However, current
  estimates of the X-ray heating of the IGM are very uncertain because
  of the poor knowledge of the abundance and spectral
  properties\footnote{Note that the mean free path of X-rays in
    primordial H--He gas is roughly proportional to the cube of the
    photon energy, so that the Universe at $z\sim 10$ is transparent
    to X-rays with $E\gtrsim 1$~keV, which diminishes the efficiency of
  ionization and heating.} of X-ray sources in the early
  Universe (see \citealt{mcquinn12} for a discusion of current
  observational constraints).

The first X-ray sources could appear in the Universe only with the
beginning of star formation. In particular, the scenario where
high-mass X-ray binaries or microquasars are responsible for early IGM
heating rests on the assumption that a significant fraction 
of Population III (Pop III) stars end their lives with a supernova (SN)
explosion, leaving behind a black hole which subsequently accretes gas
from a companion star or ambient medium and produces X-ray
emission. Alternatively, X-rays suitable for IGM heating could be
provided by primordial SNe themselves, namely by thermal emission from
the SN remnant and by inverse Compton scattering of the CMB photons on
relativistic electrons accelerated by the supernova
\citep{oh01,kityos05}. However, the efficiency of these mechanisms
(the fraction of the total SN explosion energy that goes into X-rays)
is likely to be low. 

We argue that primordial SNe can nevertheless cause a significant
preheating of the early Universe, through low-energy cosmic rays (CRs)
generated by them. A number of studies
(\citealt{broetal03,kityos05,greetal07,whaetal08}) have demonstrated
that powerful SNe associated with massive Pop III stars could readily
escape from their host minihalos into the surrounding IGM. In this
situation a significant fraction of the SN kinetic energy will be put
into low-energy CRs escaping into the IGM. These CRs could then
propagate through the Universe and gradually lose their energy by
ionizing and exciting H and He atoms, thus heating the ambient
gas. Below we demonstrate that low-energy CRs from the first SNe could
cause a significant global preheating of the primordial IGM. The
heating is expected to be somewhat (depending on the topology and
strength of IGM magnetic fields) concentrated around the Pop~III
star-forming minihalos, more so than in the case of X-ray heating,
since typical distances travelled by low-energy CRs from their sources
are larger than typical distances between minihalos but much smaller
than the Hubble horizon, whereas the mean free paths of X-rays with
$E\sim 1$~keV are comparable to the horizon.

Before proceeding, we note that the general possibility of ionization
and heating of the IGM by CRs from young galaxies was disscussed
already by \cite{ginoze65} and later by \cite{natbie93}. More 
recently, \cite{stabro07} discussed a possible impact of a uniform
cosmic-ray background on star formation \emph{within halos} in the
early Universe, considering the energy density of this background a
free parameter. Also, \cite{tueetal14} recently suggested that
CRs produced by the first generations of microquasars could contribute 
to cosmic reionization.

In our calculations, we adopt the cosmological parameters determined by
\cite{planck15}.

\section{Cosmic-ray production by primordial supernovae}
\label{s:sn}

According to the modern paradigm (see \citealt{brolar04} for a
review), the first stars appeared in significant numbers at $z\sim 20$
in dark matter minihalos of mass $\mhalo\sim \mathrm{a~few}~10^5$ --
$10^7\msun$, as a result of molecular cooling of primordial gas heated
to the halo's virial temperature $\tvir\sim 10^3$~K. Although there
have been no observations of this early epoch of star formation yet
and there remain many theoretical uncertainties and computational
difficulties, the prevailing view is that Population III was dominated
by massive and very massive, $\sim 10$--$10^3\msun$, stars (e.g.,
\citealt{hiretal14}). These stars lived for several million years and
a significant fraction or perhaps most of them eventually exploded as
powerful supernovae (SNe), with kinetic energies reaching $\sim
10^{53}$~erg \citep{hegwoo02,tometal07,hegwoo10}. This energy exceeds
by orders of magnitude the binding energy of the minihalo's gas, $\eb\sim
10^{49}$--$10^{51}$~erg (assuming that the gas contributes a fraction
$\fgas\sim 0.19$ of the mass of the halo), so a single 
SN is potentially capable of driving most of the gas from its halo and
stalling subsequent star formation in it. 

The above energy argument is, however, oversimplified, since if a SN
explodes in a sufficiently dense environment it can quickly lose a
significant fraction of its initial energy through
radiation. Nevertheless, as has been demonstrated by many authors
\citep{kitetal04,whaetal04,abeetal07,johetal07,yosetal07}, the strong
UV radiation of the Pop III progenitor star of a SN exploding in a
minihalo is expected to ionize the IGM within several kpc and almost 
completely photoevaporate gas within the virial radius of the halo,
$\rvir\sim 100$~pc, lowering the hydrogen number density there to
$\nh\sim$~0.1--1~cm$^{-3}$. As a result, the ensuing supernova
explodes in an environment almost devoid of gas, so that the SN remnant
(SNR) experiences very little radiative losses for the first $\sim
10^5$~years as it expands out to $r\sim \rvir$. At about this
distance, the SNR shock encounters increased gas density due to the
preceeding shock driven by the photoionization and, as a result,
starts to suffer significant radiation losses. Nevertheless, the SNR
retains significant energy to expand further into the HII region
created by the progenitor star and eventually disperses at $r\sim {\rm
  a~few~}\rvir$ \citep{kityos05,greetal07,whaetal08}. 

The scenario described above is only applicable to primordial SNe
exploding in minihalos, whereas the situation was likely completely
different in more massive halos, with $\mhalo\gtrsim 10^7\msun$. In
this case \citep{kityos05,whaetal08}, the progenitor Pop III star is
unable to photoevaporate the gas from the halo and the SN explodes in
a dense environment with $\nh\gtrsim 10^4$~cm$^{-3}$. As a result, the
thermal energy of the SNR is radiated away before the beginning of the
Sedov--Taylor phase and the SNR fades away within a few
pc of its origin. This is why we only consider the case of primordial
SNe exploding in minihalos in this work.

Apart from the (presumably) larger explosion energy, the case of a
primordial SNR in a minihalo at $z\sim 20$ closely resembles the more
familiar case of SNRs in the Milky Way at the present epoch. Indeed,
the gas densities encountered by a primordial SNR, $\nh\sim
$~0.1--1~cm$^{-3}$ at $r\lesssim\rvir$ and $\nh\sim
10^{-1}$--$10^{-3}$~cm$^{-3}$ at $r\gtrsim\rvir$, are similar to the
densities of the media through which SNRs propagate in the Galaxy
(e.g. $\nh\sim 0.1$~cm$^{-3}$ for Type Ia SNRs in the interstellar
medium and $\nh\sim 10^{-2}$~cm$^{-3}$ for Type Ib/c and Type IIb SNRs
in a rarefied bubble formed by the progenitor star, see, e.g.,
\citealt{ptuetal10}). Based on this similarity (although there is a
potentially important difference between the strengths of magnetic
fields in the primordial ISM and IGM and in the present-day ISM, see a
discussion in \S\ref{s:spectrum} below) it is reasonable to expect
that primordial SNRs were as efficient cosmic-ray producers as
Galactic SNRs are known to be. Moreover, we can rely on the
well-developed theory of CR production in the Galaxy in making our
estimates. 

We consider a simplistic scenario in which a single massive Pop III
star forms and then explodes in a minihalo, stalling further star
formation. In reality, it is possible that more than one massive star
will be formed within the same halo (especially if it is rather large,
$\mhalo\sim 10^7\msun$), so that several SNe will explode nearly
simultaneously. These multiple explosions can then drive a 'superwind'
that is more powerful than a single SNR (see
e.g. \citealt{schetal75,tegetal93}). On the other hand, it is possible
that a significant fraction of minihalos did not produce Pop III
stars, and hence SNe, at all, in particular due to suppression of
H$_2$ formation by a Lyman--Werner (LW) background from previous
generations of Pop III stars \citep{haietal00}. This effect should be
especially strong in overdense regions of the early Universe, as has
recently been demonstrated by \cite{xuetal13} who simulated formation
of thousands of minihalos and Pop III stars at $z=30$--15 taking into
account kinetic, thermal, chemical and radiative feedback. In
addition, there is significant uncertainty in the initial mass
function of Pop III stars and consequently in the fraction of such
stars that exploded as supernovae (rather than collapsed to black
holes) and in the energy of typical primordial SNe.

Bearing all these uncertainties in mind, we introduce two free
parameters: the average number of SNe per minihalo, $\fsn$, and the
average SN explosion energy, $\esn$. In making our estimates, we
consider the range $\esn=10^{51}$--$10^{53}$~erg, where the upper
boundary is intended to represent the case of a single
pair-instability SN.

\subsection{Cosmic-ray spectrum and energy budget}
\label{s:spectrum}

The modern paradigm of the origin of Galactic CRs builds on the
non-linear theory of diffusive shock acceleration (DSA, see
\citealt{bell13,byketal13} for recent reviews) and a detailed
treatment of the problem of CR escape from SNRs (e.g.,
\citealt{beretal96,bervoe00,ptuzir05,ptuetal10,capetal10}). According
to this theory, the time-integrated energy spectrum of CRs injected
into the ambient medium is the sum of two components: i) CRs steadily
escaping from the upstream region of the shock, and ii) CRs gradually
accumulated within the remnant and released into the surrounding
medium at the end of the SNR evolution. At any given time, the
instantaneous spectrum of the CRs escaping upstream from the shock is
peaked around a maximum momentum $\pmax(t)$, which decreases with time
over most of the evolution due to the decreasing shock velocity and
hence weakening magnetic field amplification, except for the short
initial free-expansion phase when the shock velocity, $\vsh$, is
constant. Particles with $p(t)<\pmax(t)$ cannot escape and remain
within the remnant, suffering adiabatic losses, until some later
moment when their momentum becomes larger than the current $\pmax$ at
the shock. Importantly, as emphasised by \cite{drury11}, the energy
adiabatically lost by particles within the remnant goes to driving the
shock and is thus constantly recycled into the acceleration of new
particles. 

In the time-integrated CR spectrum (see e.g. Fig.~3 in
\citealt{capetal10}), the contribution of particles leaking from the
upstream region is only important at the highest energies (near the
'knee'), whereas the rest of the spectrum is composed of particles
escaping at the end of the SNR evolution, when the remnant disperses
in the ambient medium. Interestingly, except for the knee region, the
resulting spectrum seems to be almost insensitive to the type and
environment of a Galactic supernova, with the slope of the momentum
distribution ($dN/dp\propto p^{-\alpha}$) changing from 
$\alpha\approx 3$ at $p\ll mc$  (here $m$ is the mass of the particle)
to $\alpha\approx 2$ at $p\gg mc$ (see Fig.~1 in
\citealt{ptuetal10}). In terms of particle kinetic energy, this
implies that $dN/dE\propto E^{-\beta}$ with $\beta\approx 2$ across
the entire spectrum including both the non-relativistic and
relativistic parts, i.e. the energy carried away by CRs is distributed
approximately uniformly on a logarithmic energy scale.

Recently, the \textit{Voyager 1} spacecraft has apparently left the
heliosphere modulation region and for the first time measured the
local CR spectrum down to $\sim 3$~MeV per nucleon
\citep{stoetal13}. This observed spectrum represents the result of
very strong transformation (suppression) of the CR source spectrum due
to energy losses during the diffusion of CRs through the
interstellar medium. \cite{schetal14} have demonstrated that the
\textit{Voyager 1} data are consisent with a simple model where the
solar system resides within a spatially homogeneous layer of
distributed CR sources injecting the same momentum power-law spectrum
$dN/dp\propto p^{-s}$, with $s=2.24\pm 0.12$. This corresponds to
$dN/dE\propto E^{-1.62}$ and and $dN/dE\propto E^{-2.24}$ at
non-relativistic and relativistic energies, respectively. However, as
noted by the authors, this result should currently be taken with
caution and requires further verification. It should be emphasised
that the local CR spectrum measured by \textit{Voyager 1} has 
little relation to the Galactic CR source spectrum at $E\lesssim
100$~MeV, since such subrelativistic particles essentially cannot
leave their production sites in the Galaxy, given the ISM density and 
magnetic fields (in contrast to the situation with the primordial IGM
discussed below).

The overall efficiency of CR production by SNe is not well
known. Theoretical studies of the DSA mechanism 
usually demonstrate the possibility of putting up to $\sim 50$\% of
the shock kinetic energy into acceleration of CRs. In recent
state-of-the-art simulations of DSA of ions in non-relativistic
astrophysical shocks by \cite{capspi14a}, this efficiency was found to
be $\sim 10$--20\%. On the other hand, an average CR production
efficiency of 10--30\% in SNe is required to maintain the observed CR
energy against losses from the Galaxy
(e.g. \citealt{druetal89,beretal90}).

As noted above, the case of a primordial SN exploding in a minihalo is
similar to that of SNe in the Milky Way. In particular, the evolution
of the remnant in the former case is not very different from that in
the latter, as has been demonstrated by \cite{greetal07} using
three-dimensional hydrodynamical simulations in a cosmologial setup
(their results are in good agreement with those obtained by
\citealt{kityos05} and \citealt{whaetal08}). Specifically, these
calculations show that the evolution begins with a free-expansion
phase, which continues for $\lesssim 10^4$~years and ends when the SNR
has expanded to $\sim 20$~pc. After that, a Sedov--Taylor (ST)
adiabatic phase begins, which continues for $\sim 10^5$~yr and
terminates when the shock front approaches the virial radius
$\rvir\sim 100$~pc and catches up with the previously formed 
photoheating shock associated with the progenitor. This accelerates
radiative losses and triggers a so-called snowplow phase\footnote{By
  the end of this stage additional energy losses due to inverse
  Compton scattering of the cosmic microwave background become
  important \citep{kityos05,greetal07}.}, which continues for
$10^6$--$10^7$~yr and finishes by the dispersal of the SNR at a
distance of a few~$10^2$~pc, i.e. already outside the virial radius of
the minihalo. Once the remnant has faded away, the bulk of the CRs
accelerated during its evolution get injected into the IGM. Continuing
the analogy with Galactic SNRs, most of the energy contained in the
CRs was probably generated in the ST and early snowplow phases, with
lower energy particles produced later than higher energy ones. 

An important difference between Galactic and primordial SNRs is that
the former propagate through a substantially magnetised ($B_0\sim
1$~$\mu$G) ISM, whereas the latter through a medium with likely much
weaker magnetic fields. For example, \citet{xuetal08} simulated
Pop~III star formation including seed field generation by the Biermann
battery mechanism. They found magnetic field strengths declining from
$B\sim 10^{-9}$~G in the dense protostellar core to $\sim 10^{-15}$~G
at the virial radius of the minihalo. Seed magnetic fields might also
be spewed out by stellar winds and supernova explosions of Pop~III
stars (see e.g. \citealt{bisetal73}). Once generated, seed fields of
$\sim 1$~nG can be further amplified by several orders of magnitude
via small-scale dynamo in the cores of primordial halos during their
collapse (e.g. \citealt{suretal10}).

According to the modern paradigm, the ambient magnetic field is
strongly amplified at a SNR shock during DSA of CRs. Indeed,
observations indicate that magnetic fields in young Galactic SNRs are
up to $\sim 100$ times stronger than in the ISM (see
e.g. \citealt{bell13}). Recent simulations of DSA in strong
non-relativistic shocks confirm that various instabilities can greatly
amplify the ambient field. In particular, \citet{capspi14b} find that
the amplification factor, $\bsh/B$, is proportional to the square root
of the Alfv\'{e}nic Mach number of the shock, $\ma$ (which in their
work is nearly equal to the sonic Mach number, $\ms$), for
$\ma<100$. \cite{byketal14} performed calculations reaching larger
Mach numbers, $\ma\sim 2500$, and found that magnetic field
amplification can be much stronger, $\bsh/B\propto\ma$. In fact, in
their simulations the magnetic pressure saturates at $\sim 10$\% of
the shock ram pressure. Naively this suggests that despite the
presumably much weaker ambient field in the case of a primordial SNR,
it can be amplified to the same intensity as in the case of a Galactic
SNR. However, this is based on a bold extrapolation of our current
knowledge by perhaps several orders of magnitude in $\ma$. Indeed,
taking fiducial values for the ambient number density and temperature
$\nh=10^{-2}$~cm$^{-3}$ and $T=100$~K (taking into account that the
gas upstream of the SNR is likely to be preheated by the photoionizing
radiation from the progenitor star and the SNR itself, we find the
sound speed may be $\cs=\sqrt{(5/3)kT/\mpr}\sim 1$~km~s$^{-1}$ and the
Alfv\'{e}nic speed $\va=B/\sqrt{4\pi\mpr\nh}\sim 0.02 (B/10^{-9}~{\rm
  G})$~km~s$^{-1}$. Hence, for shock velocities of $\sim
10^2$--$10^4$~km~s$^{-1}$ (corresponding to late and early stages of
SNR evolution), the sonic Mach number is likely to be $\sim
10^2$--$10^4$, similar to Galactic conditions, while the Alfv\'{e}nic
Mach number is likely to exceed $10^{4}$. Regardless of the actual
magnitude of magnetic field amplification in primoridal SNRs, it is
unlikely to significantly affect the overall efficiency of CR
production. However, the actual strength of the magnetic field should
determine the maximum energies of CRs accelerated in such shocks, as
is further dicussed below. 

As explained in \S\ref{s:loss} below, we are primarily interested in
CR protons with energies $E\lesssim 30$~MeV. The spectrum of CRs
generated by a primordial SNR probably begins at $\emin\sim
1$~keV\footnote{This corresponds to several times $\mpr\vsh^2/2$,
  where $\vsh\sim 200$~km~$^{-1}$ is the shock velocity during the
  early snowplow phase and $\mpr$ is the proton mass.} and cuts off at
$\emax\sim 10^{5}$~GeV or maybe at a much lower energy if magnetic
fields at the shocks of primordial SNRs are not amplified to the same
strength as in Galactic SNRs (since $\emax\propto\vsh\rsh\bsh$ for
DSA, where $\rsh$ is the characteristic size of the shock). Therefore,
subrelativistic CRs with $1~{\rm keV}\lesssim E \lesssim 30~{\rm
  MeV}$, which are hereafter referred to as low-energy cosmic rays
(LECRs), may provide a significant contribution, $\sim 5$--50\% to the
total CR energy budget. Here, the lower value corresponds to the
Galactic CR source spectrum inferred by \citealt{schetal14} from
\textit{Voyager~1} measurements and $\emax\sim 10^{5}$~GeV, and the
higher one to the $dN/dE\propto E^{-2}$ spectrum suggested by detailed
calculations of CR production in SNRs and $\emax\sim 10^{3}$~GeV (see
the preceeding discussion). Taken into account that the total CR
energy output is $\sim 10$--50\% of the SN explosion energy (see again
the preceeding discussion), we conclude that LECRs will carry away a
fraction $\eta\sim (0.05\div 0.5)\times (0.1\div 0.5)\sim 0.005$--0.25
of the total SN explosion energy. We may then estimate the total
energy of LECRs per minihalo as
\beq 
\elecr=\eta\esn=5\times
10^{50}\frac{\eta}{0.05}\frac{\fsn}{1}\frac{\esn}{10^{52}~\mathrm{erg}}~\mathrm{erg}.
\label{eq:elecr}
\eeq

We note that the same primordial SNRs are also expected to produce
bremsstrahlung X-ray radiation. The results of the aforementioned
simulations by \cite{whaetal08} indicate that of the order of 1\% of
the total explosion energy can be released in the form of X-rays at
the late stages of the evolution of the remnant of a primordial SN
exploding in a minihalo. The energy efficiency of X-ray
production is thus likely an order of magnitude lower than it is for
LECRs. Therefore, the X-rays from primordial SNe probably
played a less important role in heating the early Universe compared to
the cosmic rays produced by the same supernovae. 

\section{Cosmic-ray energy losses in the IGM}
\label{s:loss}

By the time of the dispersal of the supernova remnant and escape of
LECRs, the ambient gas, which was previously photoionized by the
progenitor star (and by the SNR itself, \citealt{johkho11}), could
have recombined 
and cooled down (the recombination time at $\nh\sim 
10^{-2}$--10$^{-3}$~cm$^{-3}$, the density in this region, is on the
order of a few million years or shorter). Therefore, 
the LECRs will essentially start travelling through the IGM, with the
ionization fraction $\xe\equiv n_e/(n_e+\nnh)\ge 2\times 10^{-4}$
(here the lower limit is given by the residual ionization
fraction after cosmic recombination, \citealt{zeletal69}, evaluated
for the current concordance cosmology, e.g. \citealt{loefur13}) and
lose their energy by ionizing and exciting H and He atoms. 

The loss time scale for protons in a weakly ionized medium
($\xe\lesssim 0.1$) due to ionization and excitation of hydrogen
atoms is \citep{schlickeiser02}
\beqa
\tauloss\equiv\frac{E}{dE/dt}
\nonumber\\
\approx
\left\{
\begin{array}{ll}
45\nnh^{-1}~{\rm yr}, & E< 45~{\rm keV}\\
3\times10^3\nnh^{-1}\left(\frac{E}{1~{\rm MeV}}\right)^{3/2} {\rm
  yr}, & 45~{\rm keV}\le E< 550~{\rm MeV}.
\end{array}
\right.
\label{eq:tauloss}
\eeqa
where $\nnh$ is in units of cm$^{-3}$. Note that at $E>550$~MeV,
proton energy losses are dominated by pion production, which is not
included in equation~(\ref{eq:tauloss}).    

Given the cosmic baryon fraction, $\omegab=0.05$, and the
contribution of He to the total baryon density, 25\% in mass, we
obtain the following dependence of $n_{\rm HI}$ on redshift: 
\beq
\nnh(z)=1.8\times 10^{-3}\left(\frac{1+z}{21}\right)^3 {\rm cm}^{-3}.
\label{eq:nigm}
\eeq
Substituting this expression into equation~(\ref{eq:tauloss}), we find 
\beqa
\tauloss=
\nonumber\\
\left\{
\begin{array}{ll}
  2.5\times 10^4\left(\frac{1+z}{21}\right)^{-3} {\rm yr}, & E< 45~{\rm keV}\\
  1.6\times 10^6\left(\frac{1+z}{21}\right)^{-3}\left(\frac{E}{1~{\rm
      MeV}}\right)^{3/2} {\rm yr}, & 45~{\rm keV}\le E< 550~{\rm MeV}.
\end{array}
\right.
\label{eq:tauloss_z} 
\eeqa
This time scale should be compared with the age of the
Universe at redshift $z$, 
\beq
\thub\approx 1.9\times 10^8\left(\frac{1+z}{21}\right)^{-3/2} {\rm yr}.
\label{eq:time}
\eeq
Therefore, at $z\sim 20$, only subrelativistic cosmic rays with
$E\lesssim 30$~MeV could deposit their energy into the IGM within a
time comparable to the Hubble time at that epoch. Of course, more
energetic CRs were also heating the IGM, but much less efficiently. For
example, CRs with $E\sim 300$~MeV were able to lose only $\sim 2.5$\%
of their energy within a Hubble time at $z\sim 20$.

At $E\ll 10$~MeV, one can use the following approximate formula for
the delay between CR production and ensuing IGM heating:
\beqa
\Delta z\approx
-0.75\left(\frac{1+z}{21}\right)^{5/2}\frac{\tauloss}{10^7~{\rm
    yr}}
\nonumber\\
\approx -0.1\left(\frac{1+z}{21}\right)^{-1/2}\left(\frac{E}{1~{\rm 
    MeV}}\right)^{3/2};
\label{eq:dz}
\eeqa
at $E<45$~keV, $\Delta z\approx 0$.

In reality, only some of the energy injected by LECRs into a neutral
or weakly ionized IGM will go into heat, with the rest being spent on
ionization and excitation of atoms. To evaluate the resulting
ionization and heating rates, we should first take into account that 
each direct ionization of an H atom by a CR proton is accompanied on
average by a loss of $\sim 60$~eV by the proton, with about half of
this energy going into Ly$\alpha$ line excitation, this result being
only weakly dependent of the energy of the proton
\citep{dalgri58,spisco69}. The free electron arising from ionization
of an H atom by a CR proton can ionize further H atoms. Both the 
average number of secondary ionizations and average energy released as
heat per primary ionization depend on the ionization fraction of the
gas \citep{spisco69}. If the latter is very low, for instance $\xe\sim
2\times 10^{-4}$, each free electron resulting from primary ionization of
an H atom causes on average $\phi\approx 0.75$ secondary ionizations
and the amount of energy that eventually goes into heating the gas
(via Coulomb collisions of the secondary electrons with thermal electrons)
is $\eheat\sim 8$~eV per primary ionization. Therefore, a CR proton
will deposit as heat a fraction $\fheat\sim 8~{\rm eV}/60~{\rm eV}\sim
0.13$ of its total energy. If the ionization fraction is higher, $\phi$
decreases while $\eheat$ increases, so that the heating fraction also
increases: for $\xe=0.01$, $1+\phi=1.57$, $\eheat=16$~eV, hence 
$\fheat\sim 0.27$; for $\xe=0.03$, $1+\phi=1.41$, $\eheat=22$~eV, hence
$\fheat\sim 0.37$; for $\xe=0.1$, $1+\phi=1.25$, $\eheat=26$~eV, hence
$\fheat\sim  0.43$ (see Table~1 in \citealt{spisco69}). Note that in
the case of a strongly ionized medium, with $\xe\gtrsim 0.1$, not
considered here, CR protons will release most of their energy as
heat due to direct Coulomb collisions with thermal electrons. We also
note that ionization and excitation of helium is neglected in our
treatment, as is the contribution of non-proton CRs.

Therefore, the IGM heating rate per cosmic-ray proton of energy $E$ is 
\beq
\Gamma\approx\fheat(\xe)\frac{E}{\tauloss(E)}, 
\label{eq:gamma}
\eeq
where $\tauloss(E)$ is given by equation~(\ref{eq:tauloss_z}). The
corresponding total ionization rate (number of free electrons produced
per unit time) is
\beq
\xi=[1+\phi(\xe)]\frac{E/60~{\rm eV}}{\tauloss(E)}.
\label{eq:xi}
\eeq

In order to use the above equations, we need to know the ionization
fraction $\xe$, which affects the quantities $\phi$ and and
$\fheat$. If the cosmic-ray driven mechanism discussed here were the
only reionization/heating mechanism operating at $z\sim 20$--15, we
could calculate the IGM ionization history self-consistently, by
following the evolution of $\xe$ and the resulting evolution of
$\Gamma$ and $\xi$. Such an integration should start from $\xe\sim
2\times 10^{-4}$, the minimum possible ionization degree at $z\sim 20$, 
resulting from the preceeding adiabatic expansion of the
Universe. However, as mentioned above, already at such low an
ionization level a significant fraction ($\sim 13$\%) of the energy
lost by cosmic rays goes into heating the gas, with this fraction
increasing to $\sim 27$\% as $\xe$ increases to 0.01. As the IGM ionization
fraction may increase to $\sim 1$\% due to the propagation of LECRs
and the action of some other mechanisms (e.g. X-ray heating) in
the early Universe, it is convenient to use some fiducial value for
the heating fraction, e.g. $\fheat=0.25$, fix the average number of
secondary ionizations at $\phi=0.5$ (since it is not strongly
dependent on $\xe$), and rewrite equations~(\ref{eq:gamma}) and
(\ref{eq:xi}) as follows:
\beq
\Gamma=0.25\frac{\fheat}{0.25}\frac{E}{\tauloss(E)}
\label{eq:gamma1}
\eeq
and 
\beq
\xi=\frac{E/40~{\rm eV}}{\tauloss(E)},
\label{eq:xi1}
\eeq
where $\tauloss(E)$ is again given by equation~(\ref{eq:tauloss_z}).

Therefore, only a fraction $\fheat\sim 0.25$ of the total energy
$\elecr$ released by a given minihalo in the form of LECRs
(equation~\ref{eq:elecr})  will go into heating the IGM, while the
remaining $1-\fheat\sim 0.75$ goes into ionization and excitation of
neutral atoms in the IGM. We note that the efficiency, $\fheat(\xe)$,
of CR heating discussed here is close to that of X-ray
photoionization heating \citep{shull79,fursto10}.

Using equation~(\ref{eq:elecr}), we can estimate that LECRs from a
single minihalo can ionize in total $\sim 8\times
10^{60}(\eta/0.05)\fsn(\esn/10^{52}~{\rm erg})$ H atoms in the
IGM. This number can be compared with the total number of ionizing
photons emitted by a Pop~III star over its lifetime, which is $\sim
1.5\times 10^{63}$, $\sim 7\times 10^{63}$ and $\sim 1.8\times 
10^{64}$ for $M_{\rm star}=25~\msun$, $80~\msun$ and $200~\msun$,
respectively \citep{schaerer02,kitetal04}, i.e. 2--3 orders
larger. As was discussed in \S\ref{s:sn}, these UV photons, emitted by
the progenitor star, ionize the medium within and in the vicinity of
the minihalo before the SN explosion, but cannot ionize and heat the
IGM far from minihalos.

\section{Impact of primordial supernovae on the early Universe}
\label{s:global}

To estimate the impact of LECRs produced by primordial SNe on the IGM,
we need to know the abundance of minihalos collapsed by a given
redshift. This can be estimated e.g. using the fitting formula from
\cite{sheetal01}\footnote{Specifically, we use the calculator provided
  by \citealt{muretal13}.}. The result depends on the definition of a
minihalo, i.e. on the minimum and maximum halo masses. We adopt
$\mmin=10^6\msun$ and $\mmax=10^7\msun$ as fiducial values. In the
redshift range of interest here, $z\sim 10$--30, the abundance of
minihalos is mostly sensitive to the lower boundary of the mass range;
e.g. it increases by a factor of $\sim 5$ at $z\sim 15$--20 if we use
$\mmin=3\times 10^5\msun$ instead of $\mmin=10^6\msun$. We should also
take into account that IGM heating is delayed with respect to LECR
production by $\Delta z(E)$. As discussed above (in \S\ref{s:loss}),
this delay is negligible for $E\ll 1$~MeV but becomes large for
$E>$~a~few~MeV. Taking into account the uncertainties in the LECR
spectrum and especially in the energetics of primordial SNe, it seems
reasonable to use a time lag corresponding to a fixed particle energy
$E=1$~MeV, so that $\Delta z\sim -0.1$ (see equation~{\ref{eq:dz}), in
  estimating the cumulative effect of LECRs on the IGM. 

Assuming that each minihalo produces $\fsn$ SNe with energy $\esn$,
of which a fraction $\eta$ is liberated as LECRs, and then a fraction
$\fheat$ of $\eta\esn$ is deposited as heat into the IGM, we find that
the gas in the early Universe will be heated on average by the amount
\beq
\Delta\ti(z)=\frac{2\fheat\eta\fsn\esn}{3k\fgas\rhom/\mpr}\nhalo(z+\Delta z),
\label{eq:dtemp}
\eeq
where $\nhalo$ is the comoving number density of minihalos,
$\rhom=\omegam\rhoc$ is the average (comoving) matter density in the
Universe, $\rhoc$ is the critical density, $\omegam\approx 0.3$,
$\fgas=\omegab/\omegam\approx 0.19$ is the average baryonic fraction,
and $k$ is the Boltzmann constant. In equation~(\ref{eq:dtemp}) we
have neglected the presence of helium and assumed that the IGM remains
largely neutral, so that the electrons ripped from hydrogen atoms
quickly exchange their energy with the vastly dominant neutral atoms.

The corresponding change in the IGM ionization fraction (see
equation~\ref{eq:xi1}) is
\beq
\Delta\xe(z)=\frac{\eta\fsn\esn}{40~{\rm
    eV}\cdot\fgas\rhom/\mpr}\nhalo(z+\Delta z). 
\label{eq:dxe}
\eeq

Fig.~\ref{fig:heating} shows the resulting $\Delta\ti(z)$ and $\xe(z)$
for the case where minihalos with $\mmin=10^6\msun$ and
$\mmax=10^7\msun$ form between $z=30$ and $z=10$, with $\eta=0.05$,
$\fheat=0.25$, $\fsn=1$ and three alternative values for the SN
explosion energy, $10^{51}$, $10^{52}$ and $10^{53}$~erg. In the last
case, also the result for $\mmin=3\times 10^5\msun$ is shown.

\begin{figure}
\centering
\includegraphics[width=\columnwidth]{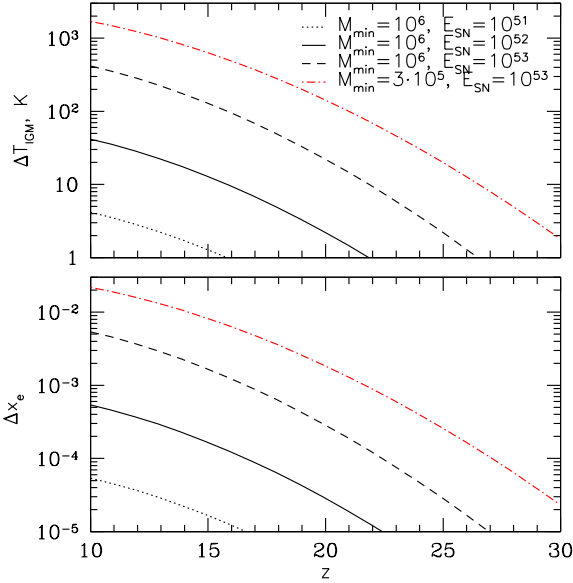}
\caption{Increment of the IGM temperature ({\sl upper panel}) and of
  the ionization fraction ({\sl lower panel}) caused by LECRs from
  primordial SNe, as a function of redshift, for three values of the
  SN explosion energy, $\esn=10^{51}$~erg (dotted), $10^{52}$~erg
  (solid) and $10^{53}$~erg (dashed). The other parameters are
  $\fsn=1$, $\mmin=10^6\msun$, $\mmax=10^7\msun$, $\eta=0.05$ and
  $\fheat=0.25$. For $\esn=10^{53}$~erg also a model with a lower
  minimum halo mass, $\mmin=3\times 10^5\msun$, is presented
  (dash-dotted).
} 
\label{fig:heating}
\end{figure}

The magnitude of the heating effect is proportinal to the assumed
average number of SNe per minihalo and is even more sensitive to the
assumed minimum mass of SN-producing minihalos. How realistic are our
fiducial parameter values $\fsn=1$ and $\mmin=10^6\msun$? To answer
this question, we may compare the number density of SNe that have
exploded by a given redshift in our simple model with the number
density of Pop~III stars formed by the same redshift in the
cosmological simulations of \citet{xuetal13}, already mentioned in
\S\ref{s:sn}. In our model there are $\sim 180$ halos with
$\mhalo=10^6$--$10^7\msun$ per (comoving) Mpc$^{3}$ formed by $z=15$,
and hence nearly the same (there is a small difference due to $\Delta
z$) number of SNe exploded by the same redshift. For comparison, in
\citet{xuetal13} simulations there are $\sim 100$ Pop~III stars and
remnants per Mpc$^{3}$ produced by $z=15$ (see their Fig.~2). Although
these numbers are rather close to each other, one should take into
account that \citet{xuetal13} considered an overdense (i.e. biased)
region of the Universe and so the number density they find for halos
with $\mhalo\gtrsim {\rm a~few}\times 10^6\msun$ is higher than the
average over the Universe by a factor of a few. On the other hand,
their simulations have a resolution of $10^6\msun$ and so may somewhat
underestimate star formation in minihalos with $\mhalo\lesssim {\rm
  a~few}\times 10^{6}\msun$, also because in more typical (i.e. less
dense) regions of the Universe suppression of star formation may be
less efficient due to the weaker LW background. Taking these
considerations into account, it appears that our $\mmin=10^6\msun$,
$\fsn=1$ model is not unrealistic. For comparison, the $\mmin=3\times
10^5\msun$, $\fsn=1$ case is probably overly optimistic since it has
$\sim 5$ times as many minihalos by $z=15$ as in the case of
$\mmin=10^6\msun$ while only a minority of such low-mass halos are
likely to be able to produce Pop~III stars and hence SNe (see again
\citealt{xuetal13} and references therein).

We conclude that, although LECRs from primordial SNe are unlikely to
significantly contribute to the reionization of the Universe
($\Delta\xe$ is less than $\sim 1$\%), they can cause a significant
early heating of the IGM. In particular, in our tentatively favoured
model with $\esn=10^{52}$~erg, $\fsn=1$ and $\mmin=10^6\msun$, the IGM
gets heated above the CMB temperature, $\tcmb=2.725(1+z)~{\rm K}\sim
30$~K by $z=12$. In the pessimistic scenario with $\esn=10^{51}$~erg, 
the resulting increment in the IGM temperature remains smaller than
$\tcmb$ over the whole considered redshift range of $z=30$--10, while
in the optimistic case of $\esn=10^{53}$~erg (corresponding to an
initial mass function dominated by massive Pop~III stars exploding as
pair-instability SNe), the IGM can be heated above the CMB
temperature already by $z\sim 18$. These estimates, however, also
depend on the $\eta$ parameter, which characterises the efficiency of
LECR production by primordial SNe and is uncertain by roughly an order
of magnitude (see the discussion prior to equation~(\ref{eq:elecr})),
as well as on the $\mmin$ and $\fsn$ parameters determining the
efficiency of Pop~III and SN production in minihalos, which are 
fairly uncertain too.

\subsection{Spatial distribution of heating}
\label{s:spatial}

If there were no magnetic fields in the IGM, the LECRs produced by
primordial SNRs would freely stream through the early Universe and
lose all of their energy within a (proper) distance $\rfree=\int
v(t)dt$, where $v(t)=[2E(t)/\mpr]^{1/2}$ is the particle velocity
which decreases from the initial value of $v_0=(2E_0/\mpr)^{1/2}$ to
zero at a rate $dv/dt= [2\mpr E(t)]^{-1/2}dE/dt$, where
$dE/dt=E/\tauloss(E)$ and $\tauloss(E)$ is given by
equation~(\ref{eq:tauloss_z}). Performing the integration, we find
\beq
\rfree~(\mathrm{proper})\approx 11
\left(\frac{1+z}{21}\right)^{-3}\left(\frac{E}{1~{\rm MeV}}\right)^{2}
     {\rm kpc}.
\label{eq:dmax}
\eeq
For $E\gtrsim 1$~MeV, this distance is larger than the average   
distance between minihalos at $z=20$, $R_0\sim 5$--10~kpc (depending
on the minimum mass of minihalos, $\mmin$). For $E\sim 30$~MeV, the
highest CR proton energy that can be deposited into the IGM within the
Hubble time at $z\sim 20$ (see Section~\ref{s:loss} above), $\rfree$
is $\sim 17$\% of the Hubble distance in that epoch. 

In the presence of magnetic fields, the LECRs will random-walk through
the IGM. In the early cosmic epochs considered here, only
very weak magnetic fields, generated during inflation or early
cosmological phase transitions, might have been present in the IGM
(see \citealt{durner13} for a review). Unfortunately, current
theoretical predictions in this respect are very uncertain. There is,
however, an interesting lower limit on the magnetic fields in voids in the
present-day Universe, determined from gamma-ray observations of
blazars \citep{nervov10,tavetal11}: $B\gtrsim 10^{-16}$~G. Taking into
account that the IGM in voids might have been contaminated by
magnetised outflows from galaxies by the present epoch but on the
other hand magnetic fields decay as $(1+z)^2$ during the expansion of
the Universe (due to magnetic flux conservation), we may adopt
$\bi=10^{-16}$~G as a fiducial value for the IGM field at $z\sim
15$--20. This of course assumes that LECRs are injected directly into
the IGM at very late stages of SNR evolution ($\gtrsim 10^6$~years
after the SN explosion, see the preceeding discussion in
\S\ref{s:spectrum}), i.e. outside the host minihalo. Nevertheless, the
SN blastwave may amplify (by the dynamo mechanism) magnetic fields in
its vicinity, which might affect the initial diffusion of LECRs
through the IGM.
 
The minimum possible diffusivity is the Bohm diffusivity, given by
$D_{B}=vr_{g}/3$, where the Larmor radius $r_{g}=\mpr
vc/eB$. Therefore, the LECRs will move away from the source by at least
\beqa
\rmf~(\mathrm{proper})\sim \sqrt{\int D_{B}(t)dt}
\nonumber\\
\sim 2.6
\left(\frac{1+z}{21}\right)^{-3/2}\left(\frac{\bi}{10^{-16}~{\rm
    G}}\right)^{-1/2}\left(\frac{E}{1~{\rm MeV}}\right)^{5/4} {\rm
  kpc}.
\eeqa
before losing all of their energy. This distance is larger than the
average distance between the minihalos at $z=20$ if $E\gtrsim
2$~MeV. In reality, IGM magnetic fields may be ordered, which would
accelerate CR diffusion compared to Bohm diffusion. Hence, the true
distances covered by LECRs are probably somewhere between $\rmf$ and
$\rfree$ (note that if $\rmf>\rfree$ then one should use $\rfree$).

Since both $\rfree$ and $\rmf$ fairly strongly depend on proton energy
and particles with $E\lesssim 1$~MeV and $1~{\rm MeV}\lesssim
E\lesssim 30~{\rm MeV}$ probably contain $\sim 2/3$ and $\sim 1/3$ of
the total energy in LECRs, respectively (assuming an intrinsic energy
distribution $dN/dE\propto E^{-2}$), the resulting IGM heating should
be somewhat concentrated towards the minihalos, especially if IGM
magnetic fields are strong enough to influence the propagation of
LECRs. Should the CR energy spectrum be steeper, $dN/dE\propto
E^{-\beta}$ with $\beta>2$, particles with $E\lesssim 1$~MeV will
carry a larger fraction of the total energy budget, leading to
stronger concentration of heating around the minihalos. In the
opposite case of a flatter CR energy distribution ($\beta<2$) the IGM
will be heated more uniformly.  

For comparison, in the case of X-ray heating, the mean free path of a
photon with energy $E$ is $\mpath=(\nnh\sigma)^{-1}$, where the
photoionization cross-section for H--He primordial gas
\citep{veretal96} can be roughly approximated as $\sigma\approx
6\times 10^{-20}(E/100~{\rm eV})^{-3.2}$~cm$^{-2}$ for $E\gtrsim
100$~eV. Therefore, taking into account equation~(\ref{eq:nigm}), an
X-ray photon releases its energy within
\beq
\rxray~(\mathrm{proper})\sim\mpath\sim
100\left(\frac{1+z}{21}\right)^{-3}\left(\frac{E}{300~{\rm
    eV}}\right)^{3.2}~{\rm kpc}.
\eeq
of the source (taking into account that the photoelectron quickly
shares its energy with the ambient IGM). Since even for
soft X-rays with $E\sim 150$~eV $\rxray$ is of the order of typical
distances between minihalos at $z\sim 20$, while for $E\gtrsim 2$~keV
$\rxray$ is larger than the Hubble distance at $z\sim 20$, X-ray
heating is likely to affect the primordial IGM more uniformly.

In reality, similarly to the case of IGM heating by soft X-ray
radiation (see e.g. \citealt{xuetal14}), LECR-driven heating should be
concentrated around high-density regions of the early Universe since
there is clustering of minihalos due to halo biasing.

\section{Discussion and conclusions}
\label{s:disc}

We have demonstrated that subrelativistic ($E\lesssim 30$~MeV) cosmic
rays produced by primordial SNe could heat the IGM to $\sim
10$--$100$~K by $z\sim 15$. The expected magnitude of
this effect is currently uncertain by at least an order of magnitude,
mainly because of our poor knowledge of the properties (in particular,
number density and initial mass function) of Pop~III stars and SNe
associated with them. Provided that there were no strong chaotic
magnetic fields in the primoridial IGM, such LECRs could deposit their
energy into the IGM at distances much larger than the virial
radii of their host minihalos ($\sim 100$~pc) but of the order of or
larger than (depending on the particle energy) typical (proper) distances
between such halos ($\sim 5$--10~kpc), and much smaller than the Hubble
horizon.

Along with heating, LECRs from primordial SNe were also ionizing the
IGM, but could not raise its ionization level by more than $\sim 1$\%
at $z\sim 15$. Although this resulting ionization fraction is 1--2
orders of magnitude higher than the initial free electron fraction
that remained after cosmic reionization, it is still too low for
recombination to start playing a significant role in the IGM at these
redshifts. Interestingly, if the IGM was already ionized to more than
a few per cent by some other mechanism (e.g. X-ray photoionization) by
that epoch, the CR heating mechanism would be operating more
efficiently, as LECRs could then release a larger fraction of their
energy as heat (more than 40\% instead of 25\% assumed as a fiducial
value in our estimates). Therefore, there might have been some synergy
between CR and X-ray heating in the early Universe. 

The overall heating effect associated with CRs from primordial SNe can
be at least comparable to that associated with X-rays from the first
generation of miniquasars and high-mass X-ray binaries. From the point
of view of 21-cm observations of the early Universe, the global
signatures of cosmic-ray IGM heating should be similar to those of
X-ray heating. In particular, depending on the cumulative power of the first
SNe,  the hydrogen spin temeperature will couple to the (increased)
IGM kinetic temperature by $z\sim 15$, which will cause the 21-cm
signal to change from absorption to emission. The spatial properties
of the 21-cm signal will depend on the topology and strength of the
IGM magnetic fields in the early Universe. Therefore, 21-cm
measurements could potentially provide information on the frequency
and energetics of primordial SNe as well as on the magnetic fields in
the primordial IGM. 

Throughout this study we assumed that LECRs are protons. In reality, a
significant fraction of cosmic rays with $E\lesssim 1$~MeV should turn
into neutral H atoms by charge exchange with ambient hydrogen (and
helium) atoms, before depositing all of their energy into the
IGM. This can be understood from a comparison of the cross-sections
for H$^+$--H charge exchange and ionization of H atoms by fast protons
(see e.g. fig.~1, based on data provided by the International Atomic
Energy Agency\footnote{http://www-amdis.iaea.org/ALADDIN/}, in the
paper by \citealt{blaetal12} devoted to a similar phenomenon occuring
during acceleration of CRs in collisionless shocks). Indeed, the
ionization cross-section is $\sim 10^4$ times larger than the
charge-exchange cross-section for $E\sim 500$~keV, with this ratio
rapidly decreasing with decreasing energy. Therefore, since each H
ionization is accompanied on average by a loss of $\sim 60$~eV by a CR
proton (see \S\ref{s:spectrum}), sub-MeV CRs will experience $\sim
10^4$ ionization and excitation collisions before losing their energy
and will thus have a significant probability to turn into H
atoms. Such neutral sub-MeV particles should be approximately as
efficient in ionizing and heating the IGM as CR protons of similar
energies, but, unaffected by IGM magnetic fields, will be able to
propagate to larger distances from their host minihalos. This is an
interesting aspect of the problem at hand, which should be considered
in more detail (in particular, taking into account helium) in future work.

Another potentially interesting implication of this study is that the
global IGM preheating caused by LECRs from the first SNe may
negatively affect further collapse of gas in minihalos and star
formation in the early Universe by raising the characteristic Jeans
mass. According to our estimates, such feedback might be important if
primordial SNe were predominantly pair-instability ones, with
$\esn\sim 10^{53}$~erg. A similar effect has been discussed for the
case of X-ray heating (see e.g. \citealt{kuhmad05}). On the other
hand, again similarly to X-rays (see e.g. \citealt{humetal14}), LECRs
from the first SNe may provide positive feedback on star formation in
neighbouring and distant minihalos by increasing the ionization
fraction in their gas and hence catalysing the formation of molecular
hydrogen, which is the main coolant in the primordial gas. 

\section*{Acknowledgments}
We thank the referee, Michael Norman, for a careful reading of the
manuscript and valuable comments. This research was supported by the
Russian Science Foundation (grant 14-12-01315). SS thanks for
hospitality the Max-Planck-Institut f\"{u}r Astrophysik, where part of
this work was done.



\begin{thebibliography}{99}

\bibitem[\protect\citeauthoryear{Abel, Wise, \&
    Bryan}{2007}]{abeetal07} Abel T., Wise J.~H., Bryan G.~L., 2007,
  ApJ, 659, L87

\bibitem[\protect\citeauthoryear{Barkana \& Loeb}{2001}]{barloe01}
  Barkana R., Loeb A., 2001, PhR, 349, 125 

\bibitem[\protect\citeauthoryear{Bell}{2013}]{bell13} Bell A.~R.,
  2013, APh, 43, 56

\bibitem[\protect\citeauthoryear{Berezhko \&
    V{\"o}lk}{2000}]{bervoe00} Berezhko E.~G., V{\"o}lk H.~J., 2000,
  A\&A, 357, 283

\bibitem[\protect\citeauthoryear{Berezhko, Elshin, \&
    Ksenofontov}{1996}]{beretal96} Berezhko E.~G., Elshin V.~K.,
  Ksenofontov L.~T., 1996, JETP, 82, 1

\bibitem[\protect\citeauthoryear{Berezinskii et al.}{1990}]{beretal90}
  Berezinskii V.~S., Bulanov S.~V., Dogiel V.~A., Ptuskin V.~S., 1990,
  Astrophysics of cosmic rays, Amsterdam: North-Holland

\bibitem[\protect\citeauthoryear{Bisnovatyi-Kogan, Ruzmaikin, \&
    Syunyaev}{1973}]{bisetal73} Bisnovatyi-Kogan G.~S., Ruzmaikin
  A.~A., Syunyaev R.~A., 1973, SvA, 17, 137 

\bibitem[\protect\citeauthoryear{Blasi et al.}{2012}]{blaetal12} 
Blasi P., Morlino G., Bandiera R., Amato E., Caprioli D., 2012, ApJ, 755, 
121 

\bibitem[\protect\citeauthoryear{Bromm \& Larson}{2004}]{brolar04}
  Bromm V., Larson R.~B., 2004, ARA\&A, 42, 79  
 
\bibitem[\protect\citeauthoryear{Bromm, Yoshida, \&
    Hernquist}{2003}]{broetal03} Bromm V., Yoshida N., Hernquist L.,
  2003, ApJ, 596, L135

\bibitem[\protect\citeauthoryear{Bykov et al.}{2013}]{byketal13} 
Bykov A.~M., Brandenburg A., Malkov M.~A., Osipov S.~M., 2013, SSRv,
178, 201 

\bibitem[\protect\citeauthoryear{Bykov et al.}{2014}]{byketal14} Bykov
  A.~M., Ellison D.~C., Osipov S.~M., Vladimirov A.~E., 2014, ApJ,
  789, 137 

\bibitem[\protect\citeauthoryear{Caprioli, Amato, \&
    Blasi}{2010}]{capetal10} Caprioli D., Amato E., Blasi P., 2010,
  APh, 33, 160

\bibitem[\protect\citeauthoryear{Caprioli \&
    Spitkovsky}{2014a}]{capspi14a} Caprioli D., Spitkovsky A., 2014a,
  ApJ, 783, 91 

\bibitem[\protect\citeauthoryear{Caprioli \&
    Spitkovsky}{2014b}]{capspi14b} Caprioli D., Spitkovsky A., 2014b,
  ApJ, 794, 46 

\bibitem[\protect\citeauthoryear{Dalgarno \&
    Griffing}{1958}]{dalgri58} Dalgarno A., Griffing G.~W., 1958,
  RSPSA, 248, 415

\bibitem[\protect\citeauthoryear{Drury}{2011}]{drury11} Drury L.~O.,
  2011, MNRAS, 415, 1807

\bibitem[\protect\citeauthoryear{Drury, Markiewicz, \&
    Voelk}{1989}]{druetal89} Drury L.~O., Markiewicz W.~J., Voelk
  H.~J., 1989, A\&A, 225, 179

\bibitem[\protect\citeauthoryear{Durrer \& Neronov}{2013}]{durner13}
  Durrer R., Neronov A., 2013, A\&ARv, 21, 62

\bibitem[\protect\citeauthoryear{Fan, Carilli, \&
    Keating}{2006}]{fanetal06} Fan X., Carilli C.~L., Keating B.,
  2006, ARA\&A, 44, 415

\bibitem[\protect\citeauthoryear{Fialkov, Barkana, \&
    Visbal}{2014}]{fiaetal14} Fialkov A., Barkana R., Visbal E., 2014,
  Natur, 506, 197

\bibitem[\protect\citeauthoryear{Furlanetto \&
    Stoever}{2010}]{fursto10} Furlanetto S.~R., Stoever S.~J., 2010,
  MNRAS, 404, 1869

\bibitem[\protect\citeauthoryear{Furlanetto, Oh, \&
    Briggs}{2006}]{furetal06} Furlanetto S.~R., Oh S.~P., Briggs
  F.~H., 2006, PhR, 433, 181  

\bibitem[\protect\citeauthoryear{Ginzburg \& Ozernoi}{1965}]{ginoze65}
  Ginzburg V.~L., Ozernoi L.~M., 1965, AZh, 42, 943

\bibitem[\protect\citeauthoryear{Greif et al.}{2007}]{greetal07} 
Greif T.~H., Johnson J.~L., Bromm V., Klessen R.~S., 2007, ApJ, 670, 1 

\bibitem[\protect\citeauthoryear{Haiman, Abel, \&
    Rees}{2000}]{haietal00} Haiman Z., Abel T., Rees M.~J., 2000, ApJ,
  534, 11 

\bibitem[\protect\citeauthoryear{Heger \& Woosley}{2002}]{hegwoo02}
  Heger A., Woosley S.~E., 2002, ApJ, 567, 532  

\bibitem[\protect\citeauthoryear{Heger \& Woosley}{2010}]{hegwoo10}
  Heger A., Woosley S.~E., 2010, ApJ, 724, 341 

\bibitem[\protect\citeauthoryear{Hirano et al.}{2014}]{hiretal14} 
Hirano S., Hosokawa T., Yoshida N., Umeda H., Omukai K., Chiaki G., Yorke 
H.~W., 2014, ApJ, 781, 60 

\bibitem[\protect\citeauthoryear{Hummel et al.}{2014}]{humetal14} 
Hummel J.~A., Stacy A., Jeon M., Oliveri A., Bromm V., 2014, arXiv, 
arXiv:1407.1847 

\bibitem[\protect\citeauthoryear{Johnson, Greif, \&
    Bromm}{2007}]{johetal07} Johnson J.~L., Greif T.~H., Bromm V.,
  2007, ApJ, 665, 85

\bibitem[\protect\citeauthoryear{Johnson \& Khochfar}{2011}]{johkho11}
  Johnson J.~L., Khochfar S., 2011, ApJ, 743, 126  

\bibitem[\protect\citeauthoryear{Kitayama et al.}{2004}]{kitetal04}
  Kitayama T., Yoshida N., Susa H., Umemura M., 2004, ApJ, 613, 631 

\bibitem[\protect\citeauthoryear{Kitayama \& Yoshida}{2005}]{kityos05}
  Kitayama T., Yoshida N., 2005, ApJ, 630, 675  

\bibitem[\protect\citeauthoryear{Kuhlen \& Madau}{2005}]{kuhmad05}
  Kuhlen M., Madau P., 2005, MNRAS, 363, 1069 

\bibitem[\protect\citeauthoryear{Loeb \& Furlanetto}{2013}]{loefur13}
  Loeb A., Furlanetto S.~R., 2013, The First Galaxies in the Universe,
  by Abraham Loeb and Steven R. Furlanetto. Princeton University Press    

\bibitem[\protect\citeauthoryear{Madau et al.}{2004}]{madetal04} 
Madau P., Rees M.~J., Volonteri M., Haardt F., Oh S.~P., 2004, ApJ, 604, 
484 

\bibitem[\protect\citeauthoryear{McQuinn}{2012}]{mcquinn12} McQuinn
  M., 2012, MNRAS, 426, 1349

\bibitem[\protect\citeauthoryear{Mesinger, Ferrara, \&
    Spiegel}{2013}]{mesetal13} Mesinger A., Ferrara A., Spiegel D.~S.,
  2013, MNRAS, 431, 621

\bibitem[\protect\citeauthoryear{Mirabel et al.}{2011}]{miretal11}
  Mirabel I.~F., Dijkstra M., Laurent P., Loeb A., Pritchard J.~R.,
  2011, A\&A, 528, AA149 

\bibitem[\protect\citeauthoryear{Morales \& Wyithe}{2010}]{morwyi10}
  Morales M.~F., Wyithe J.~S.~B., 2010, ARA\&A, 48, 127  

\bibitem[\protect\citeauthoryear{Murray, Power, \&
    Robotham}{2013}]{muretal13} Murray S.~G., Power C., Robotham
  A.~S.~G., 2013, A\&C, 3, 23

\bibitem[\protect\citeauthoryear{Nath \& Biermann}{1993}]{natbie93}
  Nath B.~B., Biermann P.~L., 1993, MNRAS, 265, 241  

\bibitem[\protect\citeauthoryear{Neronov \& Vovk}{2010}]{nervov10}
  Neronov A., Vovk I., 2010, Sci, 328, 73 

\bibitem[\protect\citeauthoryear{Oh}{2001}]{oh01} Oh S.~P., 2001, ApJ, 553, 499 
 
\bibitem[\protect\citeauthoryear{Planck Collaboration et
    al.}{2015}]{planck15} Planck Collaboration, et al., 2015, arXiv,
  arXiv:1502.01589

\bibitem[\protect\citeauthoryear{Pritchard \&
    Furlanetto}{2007}]{prifur07} Pritchard J.~R., Furlanetto S.~R.,
  2007, MNRAS, 376, 1680

\bibitem[\protect\citeauthoryear{Pritchard \& Loeb}{2010}]{priloe10}
  Pritchard J.~R., Loeb A., 2010, PhRvD, 82, 023006

\bibitem[\protect\citeauthoryear{Pritchard \& Loeb}{2012}]{priloe12}
  Pritchard J.~R., Loeb A., 2012, RPPh, 75, 086901

\bibitem[\protect\citeauthoryear{Ptuskin \&
    Zirakashvili}{2005}]{ptuzir05} Ptuskin V.~S., Zirakashvili V.~N.,
  2005, A\&A, 429, 755

\bibitem[\protect\citeauthoryear{Ptuskin, Zirakashvili, \&
    Seo}{2010}]{ptuetal10} Ptuskin V., Zirakashvili V., Seo E.-S.,
  2010, ApJ, 718, 31

\bibitem[\protect\citeauthoryear{Ricotti \& Ostriker}{2004}]{ricost04}
  Ricotti M., Ostriker J.~P., 2004, MNRAS, 352, 547  

\bibitem[\protect\citeauthoryear{Schaerer}{2002}]{schaerer02} Schaerer
  D., 2002, A\&A, 382, 28  

\bibitem[\protect\citeauthoryear{Schlickeiser}{2002}]{schlickeiser02}
  Schlickeiser R., 2002, Cosmic ray astrophysics, Berlin: Springer,

\bibitem[\protect\citeauthoryear{Schlickeiser, Webber, 
\& Kempf}{2014}]{schetal14} Schlickeiser R., Webber W.~R., Kempf A.,
  2014, ApJ, 787, 35 

\bibitem[\protect\citeauthoryear{Schwarz, Ostriker, 
\& Yahil}{1975}]{schetal75} Schwarz J., Ostriker J.~P., Yahil A.,
  1975, ApJ, 202, 1

\bibitem[\protect\citeauthoryear{Sheth, Mo, \&
    Tormen}{2001}]{sheetal01} Sheth R.~K., Mo H.~J., Tormen G., 2001,
  MNRAS, 323, 1

\bibitem[\protect\citeauthoryear{Shull}{1979}]{shull79} Shull J.~M.,
  1979, ApJ, 234, 761

\bibitem[\protect\citeauthoryear{Spitzer \& Scott}{1969}]{spisco69}
  Spitzer L., Jr., Scott E.~H., 1969, ApJ, 158, 161

\bibitem[\protect\citeauthoryear{Stacy \& Bromm}{2007}]{stabro07}
  Stacy A., Bromm V., 2007, AAS, 39, 885 

\bibitem[\protect\citeauthoryear{Stone et al.}{2013}]{stoetal13} 
Stone E.~C., Cummings A.~C., McDonald F.~B., Heikkila B.~C., Lal N.,
Webber W.~R., 2013, Sci, 341, 150 

\bibitem[\protect\citeauthoryear{Sur et al.}{2010}]{suretal10} 
Sur S., Schleicher D.~R.~G., Banerjee R., Federrath C., Klessen R.~S., 
2010, ApJ, 721, L134 

\bibitem[\protect\citeauthoryear{Tanaka, Perna, \&
    Haiman}{2012}]{tanetal12} Tanaka T., Perna R., Haiman Z., 2012,
  MNRAS, 425, 2974  

\bibitem[\protect\citeauthoryear{Tavecchio et al.}{2011}]{tavetal11}
  Tavecchio F., Ghisellini G., Bonnoli G., Foschini L., 2011, MNRAS,
  414, 3566  

\bibitem[\protect\citeauthoryear{Tegmark, Silk, \&
    Evrard}{1993}]{tegetal93} Tegmark M., Silk J., Evrard A., 1993,
  ApJ, 417, 54

\bibitem[\protect\citeauthoryear{Tominaga, Umeda, \&
    Nomoto}{2007}]{tometal07} Tominaga N., Umeda H., Nomoto K., 2007,
  ApJ, 660, 516

\bibitem[\protect\citeauthoryear{Tueros, del Valle, \&
    Romero}{2014}]{tueetal14} Tueros M., del Valle M.~V., Romero
  G.~E., 2014, A\&A, 570, LL3  

\bibitem[\protect\citeauthoryear{Venkatesan, Giroux, \&
    Shull}{2001}]{venetal01} Venkatesan A., Giroux M.~L., Shull J.~M.,
  2001, ApJ, 563, 1 

\bibitem[\protect\citeauthoryear{Verner et al.}{1996}]{veretal96} 
Verner D.~A., Ferland G.~J., Korista K.~T., Yakovlev D.~G., 1996, ApJ,
465, 487 

\bibitem[\protect\citeauthoryear{Whalen, Abel, \&
    Norman}{2004}]{whaetal04} Whalen D., Abel T., Norman M.~L., 2004,
  ApJ, 610, 14

\bibitem[\protect\citeauthoryear{Whalen et al.}{2008}]{whaetal08} 
Whalen D., van Veelen B., O'Shea B.~W., Norman M.~L., 2008, ApJ, 682,
49 

\bibitem[\protect\citeauthoryear{Xu et al.}{2008}]{xuetal08} Xu 
H., O'Shea B.~W., Collins D.~C., Norman M.~L., Li H., Li S., 2008, ApJ, 
688, L57 

\bibitem[\protect\citeauthoryear{Xu, Wise, \& Norman}{2013}]{xuetal13}
  Xu H., Wise J.~H., Norman M.~L., 2013, ApJ, 773, 83 

\bibitem[\protect\citeauthoryear{Xu et al.}{2014}]{xuetal14} Xu 
H., Ahn K., Wise J.~H., Norman M.~L., O'Shea B.~W., 2014, ApJ, 791, 110 

\bibitem[\protect\citeauthoryear{Yoshida et al.}{2007}]{yosetal07}
  Yoshida N., Oh S.~P., Kitayama T., Hernquist L., 2007, ApJ, 663, 687  

\bibitem[\protect\citeauthoryear{Zel'dovich, Kurt, \&
    Sunyaev}{1969}]{zeletal69} Zel'dovich Y.~B., Kurt V.~G., Sunyaev
  R.~A., 1969, JETP, 28, 146

\end{thebibliography}
\end{document}